\documentclass[manuscript]{aastex}

\begin{document}

\title{{\it Hubble Space Telescope\/} Expansion Parallaxes of the Planetary
Nebulae NGC 6578, NGC 6884, NGC 6891, and IC 2448\altaffilmark{1}}
\author{Stacy Palen, Bruce Balick}
\affil{University of Washington, Seattle, WA 98105}
\email{palen@astro.washington.edu, balick@astro.washington.edu}
\author{Arsen R.\ Hajian}
\affil{US Naval Observatory, Washington, DC 20392-5420}
\email{hajian@usno.navy.mil}
\author{Yervant Terzian}
\affil{Cornell University, Ithaca, NY 14853}
\email{terzian@astrosun.astro.cornell.edu}
\author{Howard E. Bond \& Nino Panagia\altaffilmark{2}}
\affil{Space Telescope Science Institute, 3700 San Martin Dr., Baltimore, MD
21218}
\email{bond@stsci.edu, panagia@stsci.edu}

\altaffiltext{1}{Based on observations with the NASA/ESA {\it Hubble Space
Telescope}, obtained at the Space Telescope Science Institute, which is
operated by AURA, Inc., under NASA contract NAS5-26555.}

\altaffiltext{2}{On assignment from the Research and Scientific Support
Department of the European Space Agency}

\begin{abstract}

We have combined two epochs of {\it Hubble Space Telescope\/} WFPC2 imaging
data with ground-based expansion velocities to determine  distances to three planetary nebulae 
(NGC~6578, NGC~6884, and IC~2448).  We used two variants of the expansion parallax 
technique---a gradient method and a magnification method---to determine the distances.  The 
results from the two methods agree to within the errors.  A fourth nebula was included in the study 
(NGC~6891), but the expansion was too small to determine the distance, and only a lower 
limit was obtained.  This is the first paper in a series which will examine at least 24 nebulae in total.

\end{abstract}


\section{Introduction}

Few planetary nebulae (PNe) in our galaxy have accurately determined
distances.  This has hampered our ability to use PNe as distance indicators in
other galaxies, as well as limiting the accuracy of derived PN properties which
are strongly dependent on distance, such as luminosity or nebular
mass.  Even the spatial distribution of PNe in the Milky Way is poorly
determined, leading to uncertainties in the total number of PNe, and the birth
rate.  Because PN progenitors are sources of carbon and nitrogen, the PN
population strongly influences the chemical evolution of the galaxy as a whole
(see, for example, Martins \& Viegas 2000).  In particular, these progenitors
contribute to the chemical evolution of carbon in the interstellar medium
(e.g., Iben 1985; Palla et al.\ 2000).  

Most PN distances have been derived statistically, and are accurate only when
taken as a whole (not for individual members of the population).  Prior to
1995, statistical distance scales relied on an assumption of a universal
physical property such as the ionized nebular mass, luminosity of the central
star, or uniform extinction along the line of sight.  These assumptions often
introduce large errors.  For example, PNe are formed from progenitor stars with
masses ranging over an order of magnitude, all of which shed enough mass to
drop to between $\sim$0.5 and $1\,M_\odot$ during the PN phase.  These
statistical scales generally gave results accurate to a factor of two at best
for individual targets (see, for example, Shklovsky 1956; O'Dell 1962; Daub
1982; Maciel \& Pottasch 1980; Cahn, Kaler, \& Stanghellini 1992).  See Terzian
(1993) for a review of the difficulties inherent in finding distances to
PNe. 

More recently, Zhang (1995) devised two additional statistical methods to
determine distances to PNe, based on empirically determined relationships
between the ionized mass and the radius of the object (mass-radius relation),
and between the radio continuum surface brightness and the radius (surface
brightness-radius relation).  Using both of these relations, he claims an
accuracy of $\sim35\%-50\%$.  These distances are still not sufficiently
accurate to enable a solution to several different PN problems.  In part, the
poor accuracy of the statistical distances is a result of calibration to a
small sample of nearby PNe with distances that have been determined by other
means.  Most recently, several efforts have been underway to improve this
nearby sample.

Trigonometric parallax has been used in recent years to find the distances to
PNe, as CCDs have improved the astrometric measurements.  Harris et al.\ 
(1997) report distances to 11 PNe accurate to 20\%, and a further 5 with
distances accurate to 50\%, all from ground-based trigonometric parallax.  The
{\it Hipparcos\/} satellite measured parallax distances to a small sample of
PNe, most to less than 50\% accuracy (Acker et al.\ 1998).  Pottasch \& Acker
(1998) compared these distances to previously determined spectroscopic
distances to three PNe, and found that the parallax distances were all much
smaller than the spectroscopic distances.   Gutierrez-Moreno et al.\ (1999)
found ground-based trigonometric parallaxes for three objects, and determined
that there was no correlation between the previously determined statistical
distances and the distances that they find from this fundamentally direct
method.  They do find, however, that distances determined from other methods,
such as expansion parallax or sodium absorption, are highly correlated with
their results.  They interpret this as a fundamental failing of the statistical
methods.  

Ciardullo et al.\ (1999) have used {\it HST\/} to discover close companions of
the central stars of PNe.  They used main-sequence photometric parallax to
derive reliable distances to nine objects with companions.  Distances to a
further three objects are given in that paper, but the association of the
binary pair is less certain. Bond \& Ciardullo (1999) also reported a
ground-based distance to NGC~246 based on photometry of the resolved binary
companion.

Expansion-parallax distances have been determined using the Very Large Array
(VLA) for eight PNe (Masson 1989a,b; Gomez, Rodriguez, \& Moran 1993; Hajian,
Terzian, \& Bignell 1993; Hajian, Terzian, \& Bignell 1995; Hajian \& Terzian
1996).  These distances are precise to roughly 10-20\% in some cases, provided
that the assumptions of elliptical morphology and constant expansion velocity
are correct.  Lower limits on about six more objects have been determined using
this method.  

The high spatial resolution offered by the {\it Hubble Space Telescope\/} ({\it
HST}) makes it potentially very useful for determining expansion parallaxes of
PNe. Prior to the work described in this paper, Reed et al.\ (1999, hereafter
Paper I) have used {\it HST\/} to determine the distance to one PN, NGC~6543,
using the expansion-parallax method .

The expansion-parallax method relies on the assumption that the expanding
nebula is a prolate ellipsoid.  Then the minor axis expansion velocity in the
plane of the sky ($V_m$) is related to the line of sight velocity ($V$) at the
position of the central star.  In the case of complex nebulae, such as
point-symmetric, or extremely bipolar nebulae, the method fails.  Fortunately,
if a nebula is close enough that $V_m$ is observable over a few years, it is
also close enough that the morphology may be accurately determined.  

We also assume, in this method, that the spectroscopy and the imaging are
sampling the same physical regimes within the object.  It is possible that the
angular expansion shows the motion of the ionization front, while the radial
velocity measures the bulk motion of the gas.  This will lead to an
underestimate of the distance to the nebula.  

A second assumption is that the nebula has been expanding at the same velocity
for the elapsed time over which the parallax is being measured.  The VLA and
{\it HST\/} have sufficient angular resolution and fidelity to measure an
expansion in just four years.  It is unlikely that the expansion speed of the
nebular gas will be dramatically altered in this short period of time.

The distance is given by the small-angle formula:

\label{exppar}
\begin{equation}
D=\frac{V_{m}}{\dot{\theta}},
\end{equation}

\noindent where $V_{m}$ is the expansion velocity of the emitting material on
the minor axis relative to the central star, and $\dot{\theta}$ is the angular
expansion of the radius of the nebula as measured on the sky.  Converting to
convenient units gives:

\begin{equation}
D_{pc}=211\frac{V_m(km/s)}{\dot{\theta}(mas/yr)}.
\end{equation}

\noindent Expansion velocities are derived from Doppler shifts along the line
of sight to the central star, corrected for geometric projection effects.  

We determined the angular expansion of the nebulae from the $V$ band {\it
HST\/} images with two techniques: the gradient method and the magnification
method, both described below.  In this paper, we report the direct measurement
of expansion parallax distances to three nebulae.  \S \ref{observations}
describes the observations and the data reduction, including the alignment of
the {\it HST\/} images.  \S \ref{modeling} describes the three dimensional
modeling technique used to determine the ratio of the line of sight velocity to
the velocity along the minor axis.  \S \ref{angexp} describes in more detail
the methods for finding the angular expansion of the nebula, and discusses the
pros and cons of each method.  In \S \ref{ngc6578}-\ref{ic2448}, we discuss the
results and derived properties for each of the four PNe.  Finally, in \S
\ref{conclusion}, we discuss possible sources of systematic errors, and future
plans for the application of the methodologies described here.

\section{Observations}
\label{observations}
\subsection{Imaging with {\it HST}: Wide Field Planetary Camera 2}

This paper describes results specifically for NGC 6578, NGC 6884, NGC 6891 and
IC 2448.  The first-epoch observations of these objects were obtained in a
$V$-band (F555W filter)  Wide Field Planetary Camera (WFPC2) snapshot survey by
Bond and R. Ciardullo (Cycle 5, GO program 6119).  The target objects were
placed in the center of the planetary camera (PC) chip.  Because this survey
was optimized for observations of central stars, it was carried out in the
broad $V$ band, rather than in a nebular emission line.  Thus, comparison to
the radial velocities obtained from [O\ III] spectroscopic data must be done
with some care.  However, since [O\ III] is almost always the main contributor
to the $V$-band flux from PNe, comparing $V$-band imaging with [O\ III]
spectroscopy will not be a problem. Other possible contributions to the $V$
band come from H$\beta$, He~I (4921, 5015, 5876, and weaker lines), and usually
negligible contributions from various Fe and N lines, as well as the nebular
continuum (which will contribute more to the $V$-band image than to the
[O~III], since the spectral window is wider). 

>From the survey of Bond and Ciardullo, we chose the 24 most spherically
symmetric objects, and reobserved many of them with WFPC2 on {\it HST\/} in
Cycles 8 and 9.  NGC 6578, NGC 6884, NGC 6891 and IC 2448 are the first four
objects with second-epoch images.  In the second epoch, we obtained [O\ III]
(F502N) and [N\ II] (F658N) data in addition to new $V$-band images.    In some
cases, the [N\ II] images allow us to investigate in more detail complementary
nebular volumes, and speculate about the nebular evolution.  Table \ref{HST
Table} summarizes the imaging observations.  

\subsubsection{Imaging Calibrations}
\label{registration}

Cosmic rays were removed using the IRAF task `crrej'.  Correction for optical
camera distortions was performed with the IRAF/STSDAS task `drizzle' (Fruchter
\& Hook 2002) with the Trauger coefficients.  After these corrections, the
chief problem was aligning images from different epochs properly.  Despite our
best efforts to repeat the first-epoch observations exactly, the images from
the two epochs were not quite perfectly aligned.  These small errors could be
due to a number of effects, such as proper motions of guidestars, or drifts in
the instruments.  Translations of fewer than five pixels in each direction and
rotations of a few tenths of a degree were required to register the two epochs
in each case. Such translations aren't important unless the optical distortion
corrections are in error; even so, they are  not likely to produce a false
positive expansion for the entire nebula.

Using IDL, we made a first order translation correction by aligning the images
relative to a ``pivot'' star.  We found the centroids, and determined the
offsets between epochs in the $x$-direction ($\delta x$) and $y$-direction
($\delta y$) by subtracting these coordinates. 

We then chose three stars in each image, found the centroids in the two epochs
([$x_1$,$y_1$] and [$x_2$,$y_2$]), and solved for the rotation required to
align the first image with the second,  
\begin{equation}
\theta_{rot}=\arctan[(y_1-\delta y)/(x_1-\delta x)]-\arctan(y_2/x_2).
\end{equation}
We compared the three values of $\theta_{rot}$ from the three stars, to check
for discrepancies due to proper motion, geometric distortions or other
complicating factors.  If the agreement was good, the three results were
averaged to determine the rotation between the two images.  If the agreement
was bad, three new stars (or a new pivot) were chosen, and the process was
repeated.

Finally, the translation of the images relative to the central star of the
nebula was determined.  For the case of an overexposed central star, we located
the star by fitting lines to each of the diffraction spikes (which make an
``X'' on the image), rather than just finding the centroid. 

To apply the transformations, the image from the first epoch was magnified by a
factor of ten, the translations were applied, and the image was de-magnified to
its original size.  (IDL allows only shifts of integer pixel size, and uses a
bilinear interpolation when magnifying, and neighborhood averaging when
de-magnifying.)  The second-epoch image was magnified, rotated, and
de-magnified.  Then the two images were subtracted and the alignment visually
inspected.  The point-spread function is rotated, and so the stars never
subtract completely.  We used the nebula itself to check the alignment, looking
for patterns of offsets which could result from a shift between one epoch and
the next.  This could result from, for example, proper motions of the
background stars.  If necessary, one image was ``nudged'' relative to the
other, with these small translations further improving the alignment.

To check that we are not introducing significant errors in this process, we
applied two tests.  First, we transformed the final images back and subtracted
from the originals.  The residuals were small in all cases (less than 4\% when
integrated across the entire nebula), and appeared mainly in places where the
nebula was faint, and close to the level of the noise in the images.  Second,
we applied the rotation to the first-epoch image, and the translation to the
second-epoch image, and subtracted the resulting difference map from the
difference map derived above.  Again, there was good agreement.  This method of
aligning the images results in only small artifacts, due mainly to the
interpolations carried out when regridding the data near under-sampled
features.

\subsection{Spectroscopy}

High spectral resolution echelle data were obtained from the four-meter
telescopes at Kitt Peak and CTIO.  In both cases, the cross-disperser was
replaced with a mirror, so that only one order, centered around [O\ III], was
obtained.  The instrument resolution was 0.06 \AA/pixel (3.7 km/s), and a
narrow-band interference filter was used ($\sim70$ \AA).  Errors in the
expansion velocity obtained from [O\ III], a strong line in all cases, are of
order 10\%.  The spectroscopic observations are summarized in Table \ref{KPNO
Table}.

\subsubsection{Spectroscopy Calibrations}

The echelle spectra were bias-subtracted and flat-fielded.  Thorium-Argon lamp
spectra were used to solve for geometric distortions by iteratively using the
IRAF tasks identify, re-identify, fitcoords and transform.  In some instances,
we used sky lines in the images to test the accuracy of these corrections. 
There were no significant deviations at the 0.1 pixel level.  These lamp
spectra were also used to calibrate the wavelength scale.   

The expansion velocity along the line of sight to the central star was found
from these data by fitting Gaussians to the double-peaked velocity profile,
subtracting the velocities of the peaks, and dividing by two. 

\section{Modeling}
\label{modeling}

The nebula is assumed to be a pinched oblate ellipsoid inclined to the line of
sight, (see Figure \ref{oblate ellipsoid}).  It is assumed to be expanding
ballistically, so that the ratio of the axes is equal to the ratio of the
velocities along those axes:
\begin{equation}
r_p/r_m=v_p/v_m, 
\end{equation}
where $r_p$ and $v_p$ are the radius and velocity along the long (polar) axis,
and $r_m$ and $v_m$ are the radius and velocity along the short (minor) axis. 
Then, the ratio of the line of sight velocity, $V$, to the true minor axis
velocity, $v_m$, is given by 
\begin{equation}
\alpha=\frac{V}{v_m}=1+(r_p/r_m-1)|\cos^n\phi| \, ,
\end{equation}
where $\phi$ is the azimuthal angle, and $n$ is functionally a shape parameter,
defining how sharply the `waist' is pinched. 

We derive $n$, $r_p/r_m$ and the inclination angle from code written in IDL
that simulates a ballistically expanding pinched-waist elliptical nebula.  Once
this nebula is created, a ``slit'' is placed across it, and a velocity spectrum
derived, similar to that observed in the echelle.  Both the projected nebula
and the spectrum are overplotted on the observed data, and the parameters are
varied manually until the shapes of the data are recovered.  Because we have
two sets of echelle data, one along the short axis, and one along the long
axis, the parameters are uniquely determined once a set of values is found
which matches the image and both spectra.  The fits to each nebula are shown in
Figures \ref{fit6578}-\ref{fit2448}, and the derived parameters are given in
Table \ref{Model Parameter Table}.  These derived parameters are used to
determine the true velocity along the minor axis from the line of sight
velocity:
\begin{equation}
v_m=\frac{V}{\alpha}.
\end{equation}  

\section{Determining the Angular Expansion}
\label{angexp}

\subsection{The Gradient Method} In the simplest case, when a nebula expands
self-similarly, a difference image (epoch~2 $-$ epoch~1) shows a bright outline
surrounding a dark outline.  When the expansion is resolved, features may
expand through several pixels, and the distance between the peak and the trough
of the difference map will give the angular expansion.  When the expansion is
unresolved, features move through less than a pixel. If the flux is high, we
can still find the angular expansion by dividing the difference in the flux
($\delta f$) by the gradient ($df/dr$), thus recovering the angular expansion
($\delta r$).  Dividing this angular expansion by the time elapsed gives an
expansion rate.  This method gives two estimates for the angular expansion, one
at either end of the minor axis.  This method is described in greater detail in
Hajian, Terzian \& Bignell (1993).

Operationally, we used IDL to make these measurements.  After choosing a cut
along the minor axis, we overplot the flux measurement along that axis, and the
derivative ($df/dr$).  Using these plots, we can find the difference in the
flux and the derivative at the location of the peak, and therefore the angular
expansion.  This is done several times, for each end of the minor axis, and the
average value is reported in Table \ref{Gradient Data}.  Errors in this method
are found from the standard deviation of these measurements.

\subsection{The Magnification Method} Following Paper I, we also determined the
angular expansion of a nebula by magnifying the first-epoch image by a factor,
{\it M}, and subtracting it from the second.  The process was repeated for a
reasonable range of {\it M} ($1<M<1.2$), to find the value of {\it M} which
minimized the residuals.  This method assumes that the entire nebula expands at
the same rate, but since we do not require the entire nebula to disappear when
we subtract the magnified epoch, but rather search for an {\it M} which gives
the best fit to the second-epoch image, the effect of this assumption is
minimized.  The angular expansion rate of the radius is given by
\begin{equation} \dot{\theta}=\frac{(M-1)\theta}{\delta T} \end{equation} where
$\theta$ is the angle between the center of the nebula and the bright rim being
used to determine the best fit, and $\delta T$ is the time elapsed between
epochs of observation.  This method shows graphically how the nebular expansion
varies from the assumed uniform expansion.  Errors are found by comparing the
difference images at various magnifications.  The largest step for which there
is no observable change between difference images is taken as the accuracy of
the measurement.  For example, if $M=1.001$ and $M=1.002$ yield
indistinguishable difference maps, but $M=1.0025$ is obviously different than
either, then the error in the measurement is 0.001.  (The error is determined
in both directions, i.e. in the preceding example, the difference map at
$M=1.0005$ is also distinguishable from the difference maps at $M=1.001$ and
$M=1.002$.  Table \ref{Magnification Data} summarizes the magnification
results.

\section{NGC 6578}
\label{ngc6578}

\subsection{Morphology}
 Figure \ref{color}  shows a color composite of the [N\ II], [O\ III] and $V$
images, while Figure \ref{fit6578}a, b and c show the $V$ band, [O\ III] and
[N\ II] images 0. All of these are images from the second epoch. In
the $V$ image, NGC 6578 exhibits a bright central core ($\theta=6\arcsec.3$),
surrounded by a fainter halo (about 11 arcseconds in diameter).  The inner core
appears to have two ``blowout'' bulbs along a single axis passing through the
central star, from slightly west of north to slightly east of south.  These
blowout lobes are co-located with halo regions of decreased brightness. 
Perhaps these are places where the inner core is not confined as tightly by the
halo as the rest of the core.  

In the [N\ II] second-epoch image, there are a few bright knots which seem to
be associated with the southeast blowout region.  Two of these knots are very
close together, and may be physically connected.  All of these knots appear to
have downstream ``tails,'' which point away from the central star, and away
from the blowout region, similar to those in the Helix nebula, but with many
fewer knots.

Along a second axis, approximately from east to west, is a much larger [N\ II]
emission axis, with the lower [N\ II] emission region containing many
approximately spherical blobs of [N\ II] emission.  None of these knots appear
to have tails of the type exhibited by the southeast knots.  This may be an
orientation effect, or may be real---there is no ``blowout'' appearance along
this axis.  The outer edge of this set of blobs is sharply bounded. The diffuse
[N\ II] emission also appears to drop to zero at approximately the same
location.  While the diffuse emission extends from the central core to the
outer edge of the blobs, the [N\ II] blobs appear to be entirely exterior to
the bright inner core.

The inner core has a bright rim, and a blotchy inner appearance, reminiscent of
a heap of soapsuds, with faint regions surrounded by bright rims.  These bright
rims appear to be outlines, not striations.  They do not in general cross each
other, but where one rim is perpendicular to another, they make a ``T''.  We
may be viewing a tightly packed bundle of smaller hollow bubbles.  When viewed
in projection, they give this blotchy appearance.  

The inner core of this nebula appears to be a waisted ellipsoid with axial
ratio 1.3:1 and an inclination angle of $25^\circ$ (see Table \ref{Model
Parameter Table}).  However, the inner core is quite asymmetric, and this is
the poorest fit of all the nebulae studied here.  

\subsection{Distance/Size Determination}

The time between observations of NGC 6578 was 4.2 years.   Figure
\ref{fit6578}d is the difference image from the epoch subtraction in $V$ band,
and frames \ref{fit6578}a, \ref{fit6578}b and \ref{fit6578}c are the $V$ band,
[O\ III] and [N\ II] images respectively.  These $V$ and [O\ III] images show a
strong correspondence, implying that the kinematic [O\ III] expansion can be
compared to the angular expansion measured in $V$.  The echelle spectrum of [O\
III] gives a line of sight expansion velocity of $19.2\pm0.5$ km/s.

The gradient method (as always, along the minor axis) yields an angular
expansion rate of $1.95\pm1.0$ mas/year, corresponding to a distance of
$1.63\pm1.07$ kpc.  The magnification method gives a magnification factor of
$1.002\pm0.0005$, which corresponds to an angular expansion rate of $1.6\pm0.4$
mas/yr, and a distance of $2.00\pm0.5$ kpc. 

\section{NGC 6884}
\label{ngc6884}

\subsection{Morphology}
	Figure \ref{fit6884} shows the image data for this object.  This nebula
has an ``s-shaped'' inner core, point-symmetric around the central star,
similar to NGC 7009.  This inner core is surrounded by a filamentary region,
perhaps a round ring inclined at an angle of approximately $45^\circ$. This is
in excellent agreement with the $40^\circ$ inclination angle derived from the
model of this nebula.  Outside this ellipse is faint [O\ III] emission, which
appears to be constrained to regions perpendicular to the ring. 

	Quite puzzling, but typical of point-symmetric bipolars, is the
presence of faint [N\ II] blobs disconnected from the inner nebula (see the red
`blobs' in Figure \ref{color}).  There are two of these, extending along an
axis quite different than any of the other symmetry axes in the nebula, at
position angle 135.  Unfortunately, our kinematic data were taken along
position angles $111^\circ$ and $200^\circ$, and so we have no kinematic
information to further investigate whether these blobs might be FLIERs.

	The inner core of this nebula is fit very well by a prolate ellipsoid
with axial ratios 1.6:1, and an inclination angle of $40^\circ$.  The nebula is
quite symmetrical, and therefore well-described by this elliptical model.

\subsection{Distance/Size Determination}
	Four years elapsed between observations of NGC 6884.  Figure
\ref{fit6884} shows the image data.  Again, the [O\ III] images are a good
match to the $V$ band data (Figure \ref{fit6884}), so we are confident that
comparing [O\ III] velocities with $V$ band expansions is reasonable. The
echelle data for this nebula indicate a line of sight expansion velocity of
$16.6\pm0.4$ km/s.

	The gradient method expansion velocity of this nebula is $1.6\pm1.0$
mas/year, which gives a distance of $1.56\pm0.98$ kpc.  Using the minor axis of
the central core to determine the magnification gives an angular expansion
velocity of $1.14\pm0.3$ mas/yr.  This corresponds to a distance of $2.2\pm0.8$
kpc.

\section{NGC 6891}
\label{ngc6891}

\subsection{Morphology}

	This nebula has an elliptical inner core (see Figure \ref{fit6891}),
with an irregular outer halo.  This is in distinct contrast with the other
three nebulae, where the outer halo is symmetrical, and the inner core is
irregularly shaped.  In detail, the inner core is fairly irregular, with
brighter filaments crossing the surface.  These filaments may have the same
structure as the filaments in NGC 6578, but they are fainter in this nebula, so
the pattern is not as clear.  There is faint [N\ II] emission beyond the [O\
III] and $V$ band emission.  This [N\ II] emission is somewhat clumpy, but it
is not possible to determine whether the clumps have tails.  There is no
evidence of blowouts or collimated outflows.

This nebula is well described as a prolate ellipsoid with axial ratio 1.5:1,
inclined about $35^\circ$.  The quite straight edges of this nebula require $n$
to be larger than for any other nebula in this study, about 6.0 fits the nebula
well.

\subsection{Distance/Size Determination}

The echelle data yields an expansion velocity of 7.67 km/s for this nebula,
which is only marginally resolved. The $V$ band and the [O\ III] images of this
nebula do not correspond well (Figure \ref{fit6891}). In the [O\ III] line, the
halo is of comparable brightness with the inner core (the two regions are
distinguished more by their morphologies than by a change in luminosity).  It
is probable that the [O\ III] spectroscopy is sampling the halo, rather than
the inner core. 

As might be expected from the low expansion velocity, the expansion was too
slow to be reliably detected by either the gradient method or the magnification
method.  However, we can place an upper limit on the expansion, of
approximately 1.6 mas/yr (the smallest expansion detected in the other
objects).  Combining this value with the marginally resolved expansion velocity
gives a lower-bound of 1.0 kpc for the distance to NGC 6891.  Correcting for
instrumental smearing, the expansion velocity may be as low as 6.5 km/s, which
slightly decreases this lower bound, giving 0.9~kpc.

\section{IC 2448}
\label{ic2448}
\subsection{Morphology}
IC 2448 is an excellent target for this project.  The nebula is elliptical and
plain, with no complicating inner and outer structure (see Figure 6).  The lack
of field stars made image alignment difficult, but the final difference image
shows a high degree of symmetry, and so we are confident that the alignment is
good.

The [N\ II] and [O\ III] emissions in this nebula are coincident, and entirely
diffuse, with no bright knots or filaments, consistent with a picture of an
old, evolved nebula.  The [O\ III] emission dominates the $V$ band images, as
is made clear by the close correspondence in morphology between the two images
(Figure 6).  We can be confident that the angular expansion observed in $V$
band is a good match to the kinematic expansion observed in [O\ III].  

The outer ring of IC 2448 appears to {\em compress} over time.  That is, it
gets thinner, with the inner radius increasing faster than the outer radius. 
This could be interpreted as an indication that a younger, faster wind is
sweeping up material, except that the brightness of the ring does not appear to
change.  A second interpretation is that the nebula is old, the central star
has ceased producing ionizing photons, and the nebula is beginning to fade.  In
the inner, denser regions, the nebula fades faster than in the outer, more
diffuse regions.  This interpretation is supported by the evolutionary age of
the nebula, which McCarthy, et al (1990) report as 8400 years.   

This nebula is very well fit by a prolate ellipsoid with an axial ratio of
1.3:1, and an inclination angle of $25^\circ$.  

\subsection{Distance/Size Determination}

Four years elapsed between observations of IC2448.  The kinematic line of sight
expansion velocity of IC 2448 is $17.9\pm0.3$ km/s.  Figure \ref{fit2448} shows
the image data.  

The gradient method yields an angular expansion velocity of $2.19\pm1.0$
mas/yr, corresponding to a distance of $1.41\pm0.64$ kpc.  The magnification
method gives an angular expansion velocity of $2.25\pm0.6$ mas/year, or a
distance of $1.38\pm0.4$ kpc. At {\it HST\/} resolution, this nebula has no
obvious variation in the proper motion with position angle.

\section{Conclusions and Future Work}
\label{conclusion}

Table \ref{ExpPar Distances} summarizes the distance determinations of the
statistical methods, as well as the expansion parallax methods under
consideration here.  In general, there is fair agreement between the three
methods, considering the large errors on the statistical distances.  IC 2448 is
the glaring exception, where the difference between the distances determined in
this paper and the statistical distances are more than a factor of two.  Martin (1994) 
found a similar discrepancy via the extinction method.  His observations put a firm upper limit of 
1.5 kpc on the distance to this nebula, since the nebula is in the foreground of a cloud
located between 1.5 and 2 kpc away.  Our measurements are consistent with this 
firm upper limit, while the statistical distances are not.

Interestingly, all of the expansion parallax distances determined here are
smaller than the statistical distances.  However, we feel that it would be a
mistake to generalize this to all nebulae.  Not only is the current sample
small, but it is also biased in favor of close nebulae.  More distant nebulae
require more time to observe an angular expansion, and also would have small
angular diameters, and are therefore unlikely to be chosen for this study.  

The magnification method and the gradient method are in good agreement in this
study.  Certainly they overlap to within the error bars.  However, we believe
there are systematics inherent in the magnification method that make the errors
much more difficult to quantify.  For example, when the nebula expands
non-uniformly, determining the best magnification is quite subjective.  For the
cases in this paper, the nebulae expand fairly uniformly, and so these errors
are small.  When two different individuals perform the magnification method on
these nebulae, their answers differ by at most $\sim$0.001.  Operationally,
this should be taken as the error in the magnification method (when propagated
through the error equations, even this error is negligible compared with the
error in measuring the size of the nebula).  In cases where the nebula departs
drastically from uniform expansion, the gradient method is more easily
quantified, and therefore gives more repeatable results.  

We have shown that the expansion-parallax method using the {\it Hubble Space
Telescope\/} gives distances to PNe.  For nebulae which are quite similar in
[O\ III] and $V$ band images, and which expand uniformly, the errors are
well-determined.  Non-uniform expansion can cause unrecognized erroneous
results, as does a discrepancy between [O\ III] and $V$ band images.  

These results are, in general, in only fair agreement with previously
determined statistical distances.  This is not much of a surprise, since these
statistical methods often use dubious assumptions.  

Future work includes an analysis of the next twenty targets, determining their
expansion parallax distances, and further constraining the systematics. 
Derived distant-dependent properties, such as the luminosity of the nebula and
central star, the density, and the mass, will be calculated in future papers. 
As more targets are analyzed, we will begin to develop a better idea of the
distribution and density of nebulae in the galaxy, improving constraints on
chemical evolution and galactic dynamics.

\acknowledgments   We would like to thank Sean Doyle for the use of his 3-D
modelling code for IDL\null. Support for this work was provided by NASA through
grant number GO7501 and GO8390 from the Space Telescope Science Institute,
which is operated by AURA, Inc., under NASA contract NAS5-26555.

\makeatletter
\def\jnl@aj{AJ}
\ifx\revtex@jnl\jnl@aj\let\tablebreak=\\\fi
\makeatother

\begin{deluxetable}{lrrrrcrrrrr}
\tablewidth{33pc}
\tablecaption{{\it Hubble Space Telescope\/} Observations}
\tablehead{
\colhead{Object}       &
\colhead{Date}         &
\colhead{Filters}      &
\colhead{Durations (secs)} &
\colhead{$\delta T$ (yrs)}}
\startdata
NGC 6578 &  8/16/1995     & $V$             &   70    &   4.2  \\
        & 10/23/1999     & $V$, [O\ III], [N\ II]  &   70, 460, 800   &        \\
NGC 6884 & 10/13/1995     & $V$             &   80    &   4.0  \\
        & 10/22/1999     & $V$, [O\ III], [N\ II]  &   80, 560, 800   &        \\
NGC 6891 & 11/20/1995     & $V$             &   3     &   3.9  \\
        & 10/22/1999     & $V$, [O\ III], [N\ II]  &   3, 360, 640   &        \\
IC 2448  & 10/10/1995     & $V$             &   16    &   4.0  \\
	& 10/ 7/1999     & $V$, [O\ III], [N\ II]  &   16, 360, 1050   &        \\
\enddata
\label{HST Table}
\end{deluxetable}

\begin{deluxetable}{lccccc}
\tablewidth{33pc}
\tablecaption{Spectroscopic Observations}
\tablehead{
\colhead{Object}       &
\colhead{Date}         &
\colhead{Location}      &
\colhead{Exp. Times (secs)} &
\colhead{$V$ (km/s)} &
\colhead{$V_{lit}$\tablenotemark{a}  (km/s)}}
\startdata
NGC 6578 & 1999     & CTIO  &   70 &  $19.2\pm0.5$ & ...   \\
NGC 6884 & 1999     & KPNO  &   80 &  $16.6\pm0.4$  & 23 \\
NGC 6891 & 1999     & KPNO  &   3  &  $7.7\pm0.8$   &   7 \\
IC 2448  & 1999     & CTIO  &   16 &  $17.9\pm0.3$    &   13.5 \\
\enddata
\tablenotetext{a}{For convenience, $(V_{lit})$ is also given for some objects.  These literature velocities are taken from Weinberger (1989), which tabulates results from several different studies.}
\label{KPNO Table}
\end{deluxetable}

\begin{deluxetable}{cccccc}
\tablewidth{33pc}
\tablecaption{Model-Derived Parameters}
\tablehead{
\colhead{Object}       &
\colhead{Position Angles}       &
\colhead{$V_p/V_m$} &
\colhead{$i$} &
\colhead{$n$} &
\colhead{$V_m$} \\
\colhead{}       &
\colhead{}       &
\colhead{} &
\colhead{($^o$)} &
\colhead{} &
\colhead{(km/s)}}
\startdata
NGC 6578 & 180, 270  & 1.3  & 25 & 0.9 & $15.1\pm0.5$  \\ 
NGC 6884 & 111, 200  & 1.6  & 40 & 1.5 & $11.86\pm0.4$ \\
NGC 6891 & 220, 310  & 1.5  & 35 & 6   & $6.70\pm0.8$  \\
IC 2448  & 135, 215  & 1.3  & 25 & 3   & $14.67\pm0.3$ \\
\enddata
\label {Model Parameter Table}
\end{deluxetable}

\begin{deluxetable}{cccccc}
\tablewidth{33pc}
\tablecaption{Magnification Data}
\tablehead{
\colhead{Object}       &
\colhead{$V_m$ }         &
\colhead{Magnification}   &
\colhead{$\theta$ }  &
\colhead{$\dot\theta$ }  &
\colhead{Distance} \\
\colhead{}       &
\colhead{(km/s)}         &
\colhead{}   &
\colhead{(mas)}  &
\colhead{(mas/yr)}  &
\colhead{(kpc)}
}
\startdata
NGC 6578 &  $15.1\pm0.5$ & $1.002\pm0.0005$ & $3359\pm90$ & $1.6\pm0.4$  & $2.00\pm0.5$\\
NGC 6884 &  $11.86\pm0.4$ & $1.002\pm0.0005$ & $2275\pm90$ & $1.14\pm0.3$ & $2.20\pm0.8$\\
NGC 6891 &  $6.7\pm0.8$  & \nodata & \nodata & \nodata  \\
IC 2448  &  $14.67\pm0.3$ & $1.002\pm0.0005$ & $4500\pm90$ & $2.25\pm0.6$ & $1.38\pm0.4$ \\
\enddata
\label{Magnification Data}
\end{deluxetable}

\begin{deluxetable}{cccccc}
\tablewidth{33pc}
\tablecaption{Gradient Data}
\tablehead{
\colhead{Object}       &
\colhead{$V_m$ }  &
\colhead{$\dot\theta$ } &
\colhead{Distance} \\
\colhead{}       &
\colhead{ (km/s)}  &
\colhead{ (mas/yr)} &
\colhead{ (kpc)}}
\startdata
NGC 6578 & $15.1\pm0.5$ & $1.95\pm1.0$ & $1.63\pm1.07$ \\
NGC 6884 & $11.9\pm0.4$ & $1.60\pm1.0$ & $1.56\pm0.98$ \\
NGC 6891 & $ 6.7\pm0.8$ & \nodata      & \nodata   \\
IC 2448  & $14.7\pm0.3$ & $2.19\pm1.0$ & $1.41\pm0.64$  \\
\enddata
\label{Gradient Data}
\end{deluxetable}

\begin{deluxetable}{ccccc}
\tablewidth{33pc}
\tablecaption{Expansion Parallax Distance Determinations}
\tablehead{
\colhead{Object}       &
\colhead{Stat. D (kpc)} &
\colhead{Mag. D (kpc)}  &
\colhead{Grad. D (kpc)}}
\startdata
NGC 6578 & 2.43\tablenotemark{1}, 2.40\tablenotemark{2}   & $2.00\pm0.5$ & $1.63\pm1.07$   \\
NGC 6884 & 2.57\tablenotemark{1}, 2.108\tablenotemark{1}, 5.4\tablenotemark{2}  & $2.20\pm0.8$ & $1.56\pm0.98$   \\
IC 2448  & 3.10\tablenotemark{2}, 3.20\tablenotemark{1}, 3.95\tablenotemark{3}  & $1.38\pm0.4$ & $1.41\pm0.64$  \\
\enddata
\tablenotetext{1}{van de Steene and Zijlstra 1994;} \tablenotetext{2}{Zhang 1995;} \tablenotetext{3}{Cahn, Stanghellini and Kaler 1992}
\label{ExpPar Distances}
\end{deluxetable}

\newpage
\figcaption{An oblate ellipsoid inclined to the line of sight: $v$ is the
observed velocity, $r_p$ is the semi-major axis, $r_m$ is the semi-minor axis,
and $i$ is the inclination angle.  \label{oblate ellipsoid}}

\figcaption{Color composites of $V$, [N\ II] and [O\ III] emission from each
nebula.  [N\ II] is red, [O\ III] is green, and $V$ is blue.  \label{color}}

\figcaption{Observations of and model fits to NGC6578: a) the original epoch 2
$V$ band image, b) [O\ III] image, c) [N\ II] image, d) difference of the two
$V$ band images taken 4.2 years apart, e) waisted ellipsoid fit to the image,
f) the fit to PA 180, d) the fit to PA 270. \label{fit6578}}

\figcaption{Observations of and model fits to NGC6884: a) the original epoch 2
$V$ band image, b) [O\ III] image, c) [N\ II] image, d) difference of the two
$V$ band images taken 4.0 years apart, e) waisted ellipsoid fit to the image,
f) the fit to PA 111, d) the fit to PA 200. \label{fit6884}}

\figcaption{Observations of and model fits to NGC6884: a) the original epoch 2
$V$ band image, b) [O\ III] image, c) [N\ II] image, d) difference of the two
$V$ band images taken 3.9 years apart, e) waisted ellipsoid fit to the image,
f) the fit to PA 220, d) the fit to PA 310. \label{fit6891}}

\figcaption{Observations of and model fits to IC2448: a) the original epoch 2
$V$ band image, b) [O\ III] image, c) [N\ II] image, d) difference of the two
$V$ band images taken 4.0 years apart, e) waisted ellipsoid fit to the image,
f) the fit to PA 135, d) the fit to PA 215. \label{fit2448}}

\end{document}